\documentclass[times, twoside]{zHenriquesLab-StyleBioRxiv-Submission}
\usepackage{blindtext}
\usepackage{graphicx}
\usepackage{dblfloatfix} 
\usepackage{lineno}
\usepackage{comment}
\usepackage[normalem]{ulem}
\usepackage{color, colortbl}
\usepackage{MnSymbol}

\usepackage{mathrsfs}
\usepackage{dblfloatfix}
\usepackage{booktabs}
\leadauthor{Taylor} 

\begin{document}

\renewcommand{\thesubsection}{\thesection\Alph{subsection}}
\vspace{6mm}
\title{\textcolor{black}{Fast 360° 3D Metrology for  \\ Directed Energy Deposition}}

\shorttitle{Fast 360° 3D Metrology for DED}

\author[1]{James Taylor}
\author[1, *]{Jiazhang Wang}
\author[2]{Guanzhong Hu}
\author[2]{Zihan Chen}
\author[2]{Rowan Rolark}
\author[2]{\\ Jian Cao}  
\author[2]{Ping Guo} 
\author[1, *]{Florian Willomitzer}

\affil[1]{Wyant College of Optical Sciences, University of Arizona, Tucson, AZ, USA}
\affil[2]{Department of Mechanical Engineering, Northwestern University, Evanston, IL, USA}

\affil[*]{Correspondence: jiazhangwang@arizona.edu ; fwillomitzer@arizona.edu}

\maketitle

\vspace{5mm}

\section*{Abstract}
\vspace{1mm}
Directed Energy Deposition (DED) is a metal additive manufacturing process capable of building and repairing large, complex metal parts from a wide range of alloys. Its flexibility makes it attractive for multiple industrial applications, e.g., in  aerospace, automotive and biomedical fields. 
However, errors or defects introduced at any stage of the printing process can, if undetected, significantly impact the final result, rendering the printed part unusable. Potential in-situ methods for correcting printing defects require fast and high-resolution on-the-fly 3D inspection inside the machine, but existing 3D imaging techniques often lack full surface coverage, require bulky setups, or operate too slowly for layer-wise feedback.
In this paper, we present a fast, multi-view polarized fringe projection profilometry (FPP) system designed for real-time 3D inspection during DED printing. Multiple camera-projector pairs are arranged around the deposition area to measure the 3D surface in a single shot from different viewpoints, while cross-polarized image filtering suppresses specular reflections caused by varying surface reflectance across different alloys. The final 360° reconstruction is obtained via joint registration of the captured multi-view measurements. Our  prototype has been deployed inside a DED system, and our first experiments demonstrate a depth precision  of $\delta z < 55 \mu m$ and a depth accuracy of $64\mu m$, both evaluated on different partially reflective and ``shiny'' metal surfaces, enabling high-quality, layer-wise monitoring for closed-loop DED control.

\section{Introduction} 
\vspace{1mm}
Directed Energy Deposition (DED) \cite{SVETLIZKY2021271,dass2019state,feenstra2021critical} is a powerful metal additive manufacturing process that uses a focused energy source, typically a high-power laser, to create a melt pool into which metallic powder or wire is continuously fed. This forms beads that are stacked layer by layer to fabricate parts with complex geometries and tailored microstructures. Unlike powder bed fusion (PBF) methods \cite{king2015laser,singh2020powder}, DED is open-architecture, highly scalable, and capable of depositing material onto existing components, making it particularly suited for larger builds, multi-material fabrication, and on-the-fly repair applications in aerospace, energy, automotive, and tooling industries~\cite{SVETLIZKY2021271,app9163316}.
\vspace{3mm}

The nascent field of DED printing and processing introduces significant new metrology challenges that differ fundamentally from those in PBF, where the powder bed provides a relatively uniform, diffuse surface that can be easily measured by conventional optical metrology methods. In contrast, DED surfaces typically show mixed surface reflectance properties and have a ``shiny'' appearance. These surface types are particularly hard to measure for common optical 3D sensors, as most sensor principles are fundamentally limited \cite{hausler2011limitations,hausler2022reflections} to either purely diffuse \cite{xu2020status} or purely specular surfaces \cite{wang20253d,knauer2004phase,wang2025accurate}. Moreover, the DED-produced surface shapes are rough, irregular, and highly dynamic due to localized heating and rapid solidification, all of which can lead to defects such as porosity, lack of fusion, and geometric distortion. These defects may form unpredictably and accumulate over time, underscoring the need for in-situ layer-wise geometric monitoring. Several optical monitoring strategies have attempted to address this need \cite{article2,Chen02012021,hu2024digital}. For instance, the authors of \cite{article2} utilized a CCD camera to observe the side profile of the growing build and adjusted the deposition height based on the silhouette contours. While this approach enables basic feedback control, it is fundamentally limited to two-dimensional profiles. The complex material interactions and rapidly evolving topography of DED, however, often result in steep slopes, deep melt tracks, and complex surfaces, requiring a ``full'' 3D measurement for process control.

\vspace{0.5mm}
The authors of \cite{Chen02012021} later introduced a laser line scanner to evaluate the part's surface topography during DED, demonstrating the ability to detect surface deviations. However, single-line scanners inherently suffer from low 3D data density and require a multitude of camera images. For instance, obtaining a  ``full'' or ``pixel-dense'' 3D model with a 1 Megapixel camera  ($1000pix \times 1000 pix$) would require about 1000 camera images, which need to be sequentially captured while the line is scanned over the object \cite{willomitzer2019single,willomitzer2017single}. 
The 3D imaging literature also offers other established 3D imaging methods to tackle the problem: Passive stereo vision \cite{marr1979computational} and time-of-flight imaging \cite{hansard2012time, ballester2024single} are widely used due to their flexibility and ease of deployment; however, their depth measurement precision is often insufficient for resolving the subtle layer-wise geometry changes relevant to DED process monitoring. Shape-from-polarization \cite{atkinson2006recovery,wang20253d, wang2026physics} can handle mixed diffuse-specular surfaces, but typically requires strong surface priors and suffers from inherent ambiguity issues. 
\vspace{1mm}
High-precision optical metrology techniques such as interferometry \cite{steel1985interferometry} and deflectometry \cite{knauer2004phase,wang2025accurate} can provide excellent data quality, but are either impractical for in-situ deployment or do not perform well if the encountered surface still has residual roughness.

\vspace{1mm}

Fringe Projection Profilometry (FPP) \cite{Srinivasan:84, xu2020status,zuo2018phase} is typically the gold standard for measuring dense and highly accurate 3D data of diffuse surfaces, but the conventional temporal phase-shifting approach of fringe patterns also requires the capture of a temporal sequence of fringe images. This means that for both single laser-line scanning and conventional multi-shot phase-shifting FPP, the object is not allowed to move or change during the measurement period, which significantly increases the total DED printing time and system complexity, and makes related approaches unsuitable for fast in-process monitoring where thermal dynamics evolve rapidly. Moreover, most standard active triangulation techniques (like FPP or line triangulation) are, as discussed,  not well suited to measure shiny metallic surfaces without further modification, and are prone to artifacts caused by specular reflection highlights. Effective geometric monitoring of DED, therefore, demands a method that can handle surfaces with varying reflectance properties, combined with high-speed acquisition.

\vspace{1mm}

Another important factor for effective DED monitoring is achieving comprehensive 360° coverage of the printed part simultaneously across all viewing angles, which, to our knowledge, has not been demonstrated in prior work.  Instead, current monitoring approaches are limited to one specific viewing direction, which is, in many cases, insufficient for DED, where steep sidewalls, overhangs, and the presence of the deposition head frequently cause occlusions that hide critical regions. 
Achieving the required comprehensive coverage is technically challenging: the confined chamber volume restricts sensor placement; illumination and imaging hardware must be carefully calibrated across multiple viewpoints; and residual specular reflections and registration errors can easily degrade data quality. Overcoming these barriers requires both hardware optimization and algorithmic strategies for calibration, synchronization, and robust fusion of overlapping views.

\vspace{1mm}

In this paper, we introduce a full 360° 3D measurement system that employs a single-shot FPP strategy based on Fourier Transform Profilometry (FTP) \cite{takeda1983fourier,su2001fourier}. FTP delivers dense 3D measurements, but unlike traditional phase-shifted FPP, it requires the capture of only one single camera image to calculate one 3D image, which significantly speeds up the acquisition time. To measure mixed reflectance metallic surfaces with high precision, we apply cross-polarization filtering that filters out unwanted specular lobes. Our system integrates multiple camera–projector pairs distributed 360° around the build plate, each acquiring the surface from a distinct perspective. The measurements are then jointly registered into a unified 3D reconstruction, enabling synchronized, full-field capture from all viewpoints. This architecture supports high-throughput, layer-wise geometric inspection with precision,  compatible with the real-time requirements of in-situ DED processes.
We summarize our contributions as follows:
\vspace{2mm}

\begin{itemize}
    \item A compact, multi-view, single-shot fringe projection profilometry (FPP) system designed specifically for 360° in-situ DED monitoring. 
    \vspace{0.5mm}
    \item Integration of high-speed depth acquisition, polarization-based reflectance control, and a registration pipeline that selectively fuses multiple views via normal-based correspondence filtering. 
    \vspace{0.5mm}
    \item Validation on complex metallic parts inside a production DED chamber with $64\mu m$ depth reconstruction accuracy, as well as on designated test objects demonstrating a depth precision $\delta z < 55 \mu m$ in a single shot. 
    \vspace{-2.8mm}
    \item First steps towards a scalable foundation for next-generation, closed-loop 360° metrology in DED additive manufacturing.
\end{itemize} 

\vspace{1mm}
\section{\textcolor{black}{Methods}} 
\label{sec:Methods}
\vspace{1mm}
\begin{figure}[t]
\centering
\includegraphics[width=\columnwidth]{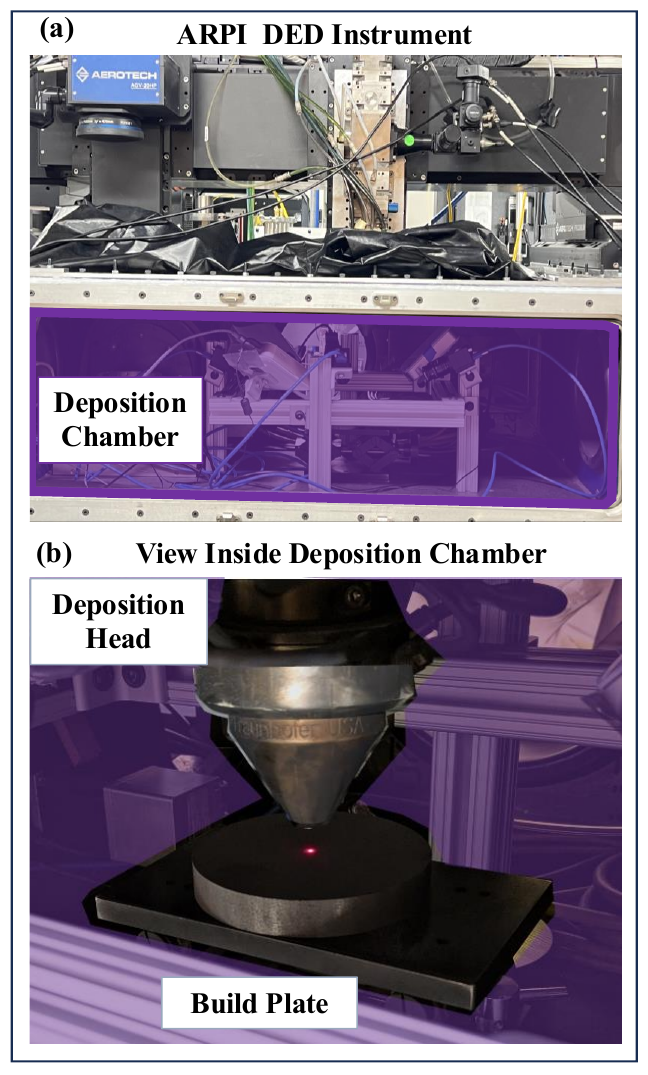}
\caption{Directed Energy Deposition (DED) chamber and hardware. 
(a)~The full ARPI system with the deposition chamber highlighted in {purple}, indicating the rough dimensions ($\sim 650mm \times 500mm \times 300mm$) into which our multi-view measurement system must fit. 
(b) Interior view of the chamber showing the two key components, deposition head and build plate, relevant to both DED functionality and the design of our multi-view system.}
\label{fig:DED_chamber}
\end{figure}

We introduce  an in-situ, 360° 3D imaging system for precise and fast surface measurement in DED. The system is based on Fringe Projection Profilometry (FPP) and is specifically tailored to meet the unique challenges of measuring metallic object surfaces in DED environments. To this end, our methods integrate multiple imaging and reconstruction techniques, including single-shot phase evaluation, joint system calibration, 3D feature-based multi-view surface registration, and polarization filtering. The following subsections provide a detailed description of each component. 

\vspace{0.5mm}
\subsection{\textcolor{black}{DED principle and hardware}} 
\label{subsec:DED}
The DED hardware used in this paper is the Additive Rapid Prototyping Instrument (ARPI) \cite{ARPI}, a laser powder-blown DED system (see Fig.~\ref{fig:DED_chamber}). The ARPI deposition chamber ($\sim 650mm \times 500mm \times 300mm$) accommodates two critical components: the deposition head and the build plate (see Fig.~\ref{fig:DED_chamber}(b)). The deposition head integrates a high-power laser with a coaxial powder delivery system, ensuring uniform melting and deposition. Powder flow is carefully regulated through adjustable gas-driven nozzles, enabling precise control over deposition rates and geometric accuracy. The build plate serves as the substrate upon which material deposition occurs. Precise control of the deposition head's translation, rotation, and overall trajectory allows complex geometries to be fabricated layer by layer. This integrated DED system provides robust control over the deposition process, enabling the production of metallic parts with tailored microstructures and optimized geometrical features for demanding industrial and research applications.

\subsection{\textcolor{black}{In-situ 360° FPP system and calibration}} 
\label{subsec:FPPcalibration}

\begin{figure}[b]
    \centering
    \includegraphics[width=1\columnwidth]{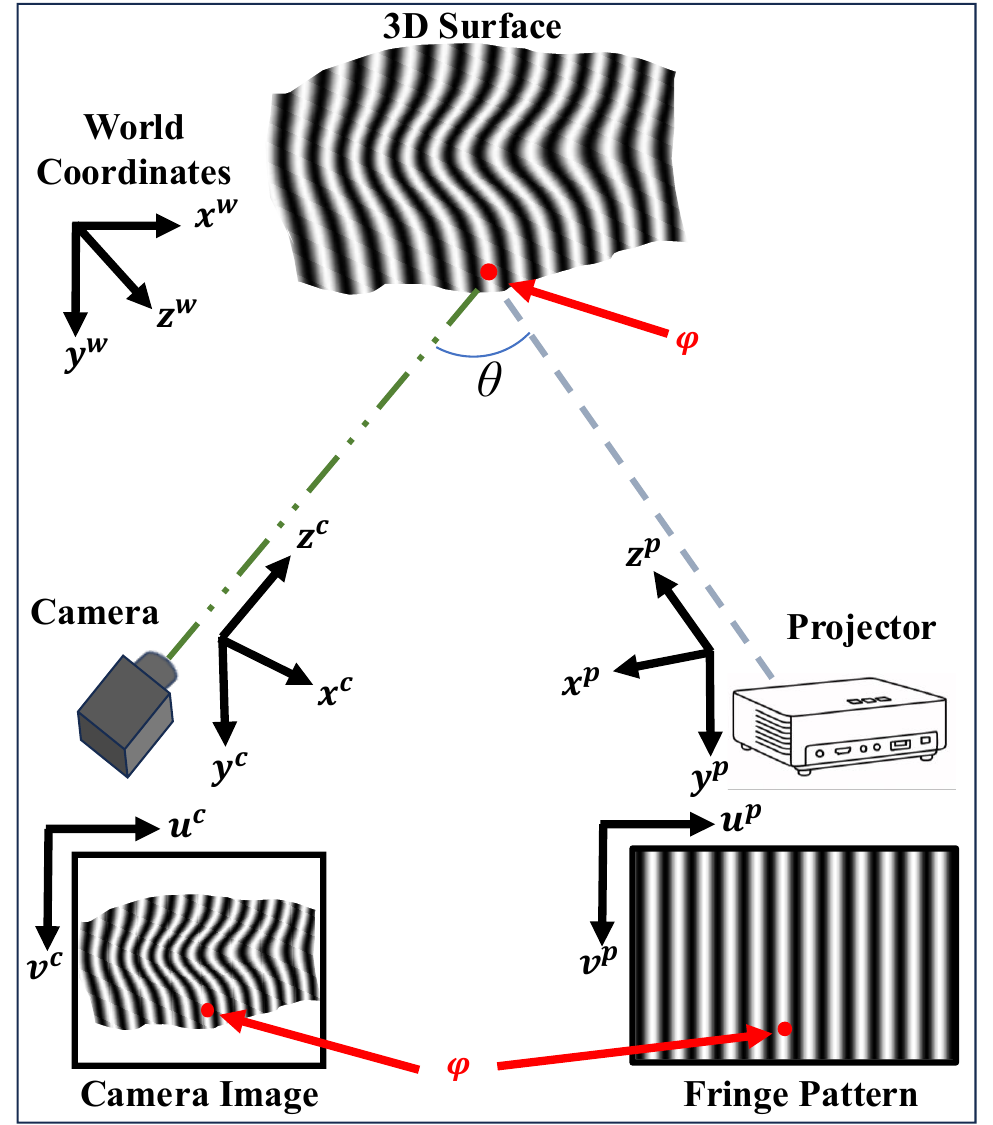}
    \caption{Schematic of a calibrated FPP system, where a camera images the deformation of a projected fringe pattern. A representative point on the surface (shown in red) illustrates how the phase information $\varphi$ of the scene establishes the correspondence between camera \((u^c, v^c)\) and projector \((u^p, v^p)\) pixels. From this correspondence, the known geometry between the projector and the camera can be used to triangulate the depth of the scene.}
    \label{fig:tri_schematic}
\end{figure}

Fringe Projection Profilometry (FPP) is a widely used optical metrology technique that enables accurate and dense optical 3D surface reconstruction by analyzing the deformation of light patterns. A calibrated projector projects a known sinusoidal fringe pattern onto the object surface, which is observed with a camera under a triangulation angle $\theta$. Due to the shape variations of the object surface, the fringes appear deformed in the camera image (see Fig. \ref{fig:tri_schematic}). This deformation, quantified as a variation in the local phase $\varphi$ of the observed pattern, is used to calculate the depth profile of the object. To relate the captured fringe deformation to the actual 3D geometry of the measured surface, a phase-correspondence must be established between each camera and projector pixel. Obtaining two corresponding points in the camera image and the projector image then allows for 3D point reconstruction via triangulation (see Fig.~\ref{fig:tri_schematic}). In conventional systems, the fringe phase in each pixel (which establishes correspondence) is typically obtained via temporal phase-shifting techniques \cite{Srinivasan:84,zuo2018phase}, where the projector sequentially displays several phase-shifted fringe patterns. One of the most common approaches is the so-called four-step phase-shifting algorithm \cite{Srinivasan:84}, in which four fringe images with relative phase shifts of \(\pi/2\) are utilized. As these multi-shot techniques are not used to capture 3D data in our system, we refer to \cite{willomitzer2019single,FENG2021106622,liu2010dual,Salahieh:14, willomitzer2020hand} for more information about phase shifting FPP.

\begin{figure}[t]
    \centering
    \includegraphics[width=1\linewidth]{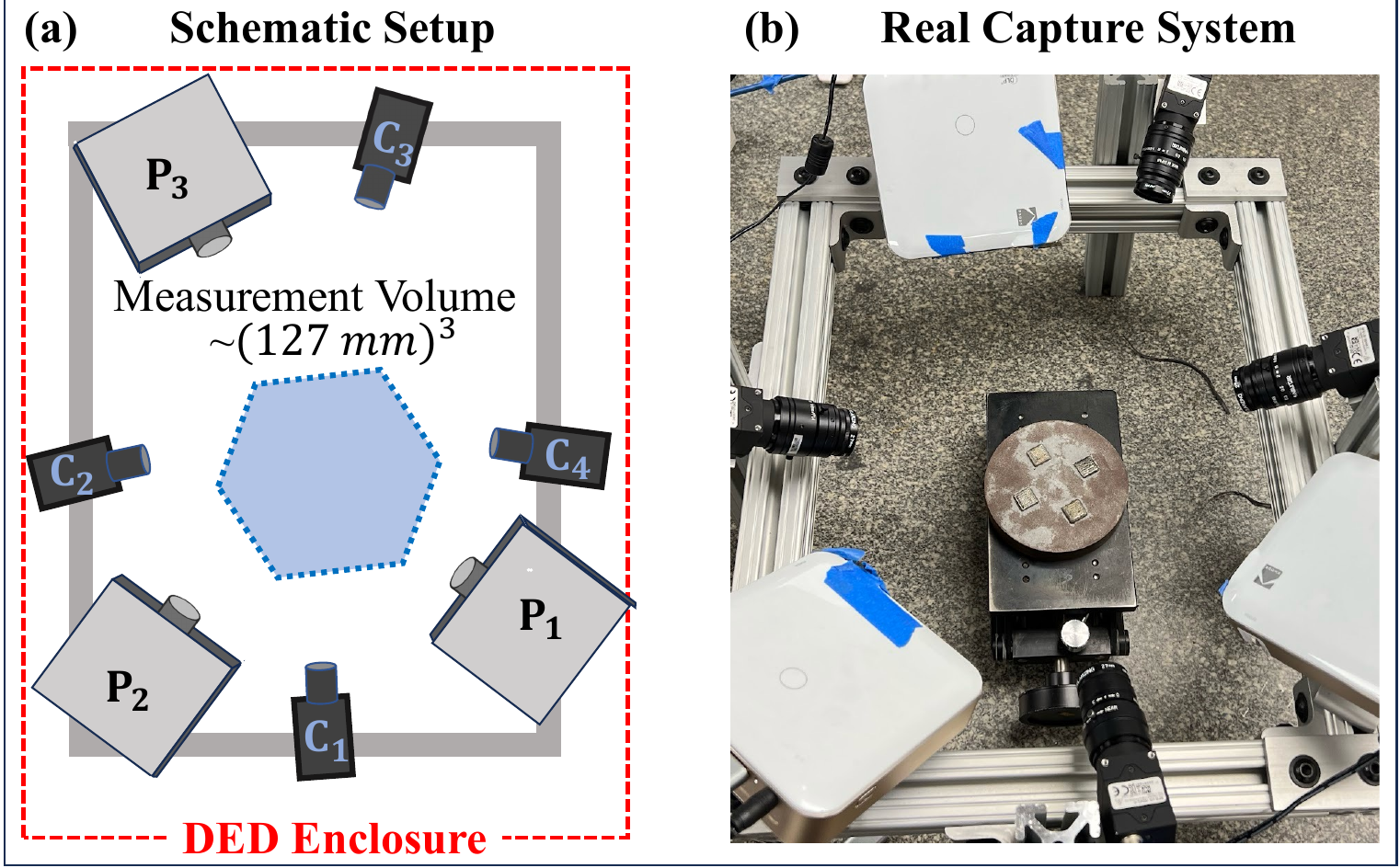}
    \caption{360° multi-view system. (a) The schematic image of the system shows the arrangement of multiple FPP sub-sensors around the defined measurement volume. While fitting inside the DED deposition chamber, the cameras and projectors are also positioned to avoid obstructing any of the ARPI internal mechanisms.(b)Image of the real capture system placed outside the DED chamber.}
    \label{fig:layout}
\end{figure}

\begin{figure*}[t]
\centering
\includegraphics[width= \linewidth]{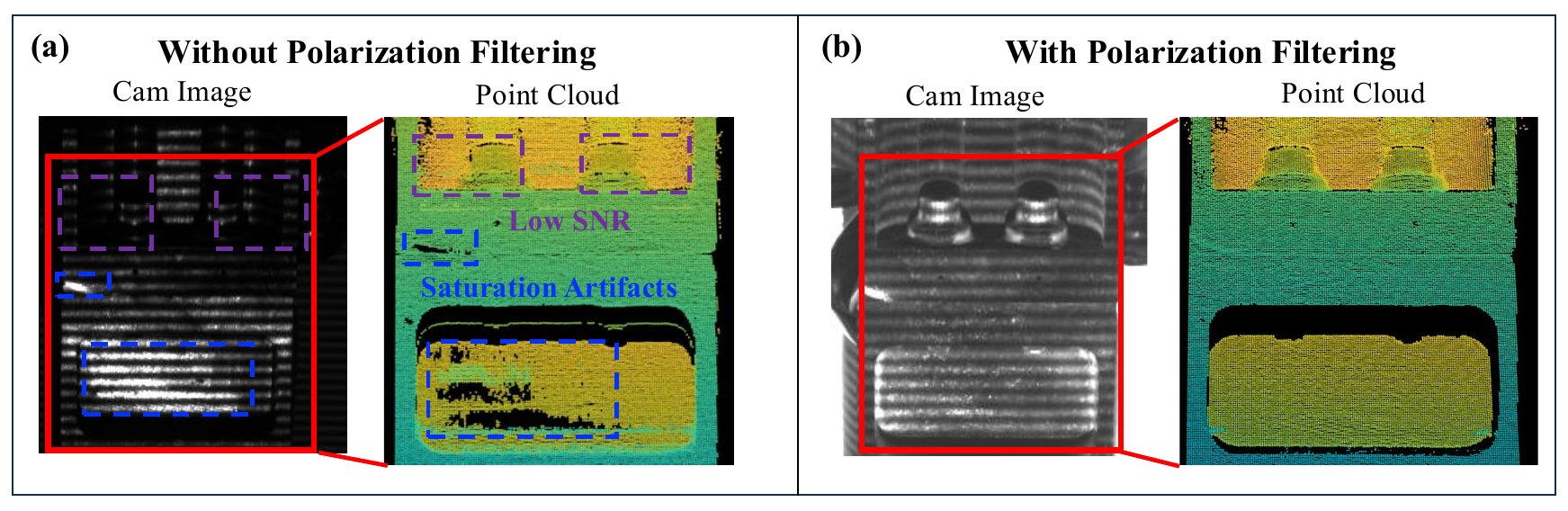}
\vspace{-1.5mm}
\caption{Specular-diffuse separation via polarization filtering. (a) Example measurement without polarization filtering. The camera image frame shows a visibly degraded fringe pattern signal, such as strong saturation due to specular reflection, and regions with low SNR. The corresponding regions in the evaluated 3D point cloud show distinct artifacts or strong noise. (b) Same measurement with specular reflections suppressed by polarization filtering. The fringe pattern is homogeneously  visible in the image frame, with a better signal throughout the measurement surface. }
\label{fig:polarization_filtering}
\end{figure*}

\vspace{1mm}
In contrast to standard FPP systems, our proposed 360° FPP system uses a fast single-shot acquisition approach (described in sec. \ref{subsec:FTP}) and consists of multiple camera-projector pairs, each representing an independent FPP sub-sensor.  Our current prototype system consists of four monochrome cameras and three digital projectors arranged around the DED build plate (see Fig.~\ref{fig:layout}). Although this results in theoretically 12 independent FPP sub-sensors, the actual number of used pairs varies with application and printing geometry. For most examples shown in this paper, we utilized 3-4 FPP sub-sensors. Each pair is modeled and calibrated using the pinhole camera approximation: A 3D point is defined as \(\mathbf{X}^w = [x^w, y^w, z^w, 1]^T\) in world coordinates (see Fig.~\ref{fig:tri_schematic}), and is mapped onto the respective camera and projector chip (image plane) via their respective intrinsic and extrinsic parameters.

\vspace{1mm}
We first calibrate each camera independently using a planar checkerboard target, following Zhang’s method \cite{888718}. Multiple images of the calibration board are acquired at varied poses inside the measurement volume, and sub-pixel precise corner locations are extracted. The camera intrinsics \( \mathbf{K} \) and lens distortion coefficients are estimated using a nonlinear optimization procedure, which minimizes the reprojection error between observed and projected checker corners across all frames \cite{Heikkila1997}. 
The projectors in our system are calibrated in the same way, i.e., by treating each projector as an inverse camera \cite{moreno2012simple}. Since a projector cannot directly observe the calibration board, we project phase-shifted sinusoids onto the calibration board to establish correspondence between the camera and the projector for each checkerboard position. Eventually, we can obtain the exact projector pixel coordinate for each observed checker corner, effectively allowing the projector to ``see'' the board through the calibrated camera.
Given this procedure, we estimate the intrinsic matrix \( \mathbf{K} \) and extrinsic pose \( [\mathbf{R} \,|\, \mathbf{T}] \) of each projector in each FPP sub-sensor by minimizing the reprojection error, analogous to the camera calibration procedure.

\vspace{1mm}
After calibrating each camera and projector individually, the final step is to establish a common reference frame across all four cameras. This requires estimating the relative poses (extrinsics) of the cameras with respect to each other. To this end, we capture multiple images of the same planar checkerboard at varied poses, ensuring that all four cameras observe the calibration board simultaneously. Sub-pixel accurate corner detections are extracted in each view and used to constrain the camera poses into a shared coordinate system.
Eventually,  the extrinsic parameters---namely the rotation $\mathbf{R}_c$ and translation $\mathbf{T}_c$ of each camera---are jointly optimized to minimize the reprojection error of all checkerboard corners across all  FPP sub-sensors \cite{PEREZ2020103256}.  After choosing  Camera~1's coordinates as the joint ``reference frame'', the multi-view system is calibrated to reconstruct 360° 3D scenes.

\subsection{\textcolor{black}{Specular-diffuse separation via multi-view polarization filtering}} 
\label{subsec:specularfiltering}

The ``shiny'' reflectance properties of metallic surfaces, common with DED, present a significant challenge for most active triangulation principles (including FPP), as these methods are designed for matte Lambertian surfaces \cite{dashpute2023event,hausler2022reflections}. When imaging glossy metallic surfaces, extended specular lobes from directed reflections can saturate image sensors and introduce large errors and artifacts in the captured 3D point cloud \cite{Zhang2023HighDynamicRange}. To mitigate this effect and mainly capture the diffuse part of the light signal, which is backscattered from the metallic surface, we apply cross-polarization filtering \cite{chen2007polarization}. As specular reflection preserves the polarization state of the incident light, while diffuse scattering scrambles the polarization state, polarization filtering can be employed to selectively suppress the unwanted specular component. Consequently, we add a linear polarizer in front of each projector to define the polarization state of the emitted light. The corresponding camera in each FPP sub-sensor is then equipped with a second linear polarizer,  oriented exactly to filter out the directly reflected light that maintains its polarization state when reflected off the metallic surface.

\vspace{1mm}
To demonstrate and evaluate the effect of polarization filtering on our measurements, we performed a controlled comparison on a shiny metallic test part under identical illumination and viewing geometry. Figure~\ref{fig:polarization_filtering}(a) shows a raw fringe image acquired without polarization filtering and the corresponding 3D reconstruction. Several regions exhibit strong specular highlights, with saturated pixel values that corrupt the underlying sinusoidal pattern information. As expected, these saturated zones lead to significant artifacts in the resulting 3D reconstruction. Moreover, due to the large dynamic range of the scene, some image regions are too dark to obtain a signal with sufficient SNR, leading to increased noise in the respective 3D reconstruction. In contrast, Fig.~\ref{fig:polarization_filtering}(b) shows the corresponding fringe image acquired under the cross-polarized configuration. It can be seen that the majority of specular glare is effectively suppressed, and high SNR fringe visibility is preserved across  a large area of the object surface. The resulting 3D reconstruction demonstrates significantly improved continuity and surface fidelity, with no visible artifacts.

\vspace{1mm}
We note that our multi-view arrangement can further improve this approach by mitigating the few remaining surface points where saturation occurs. Some specular reflections can still be present after polarization filtering. This is because the cameras' filters only remove light whose polarization is along a specific axis, which is most likely to block light from direct reflection. Accordingly, incident polarized light along a different axis may still arrive on the sensor. This can occur as a result of inter-reflections or complex surface geometry, and the specular component from this reflection is not effectively suppressed. Because these ``unfavorable'' reflections are dependent on viewing angle, they are likely not   present for the same surface point observed from a different camera position. So if there are remaining sparse saturated regions for one camera, a camera at a different position with an overlapping view will observe an uncorrupted signal from this surface area. During the multi-view registration and normal-based fusion process discussed in sec. \ref{subsec:registrationprocess}, the corrupted 3D data is then replaced by the artifact-free data from the other viewpoint.

\subsection{\textcolor{black}{Single-shot FPP measurements}}
\label{subsec:FTP}
\begin{figure}[b]
\vspace{-3mm}
\centering
\includegraphics[width=\linewidth]{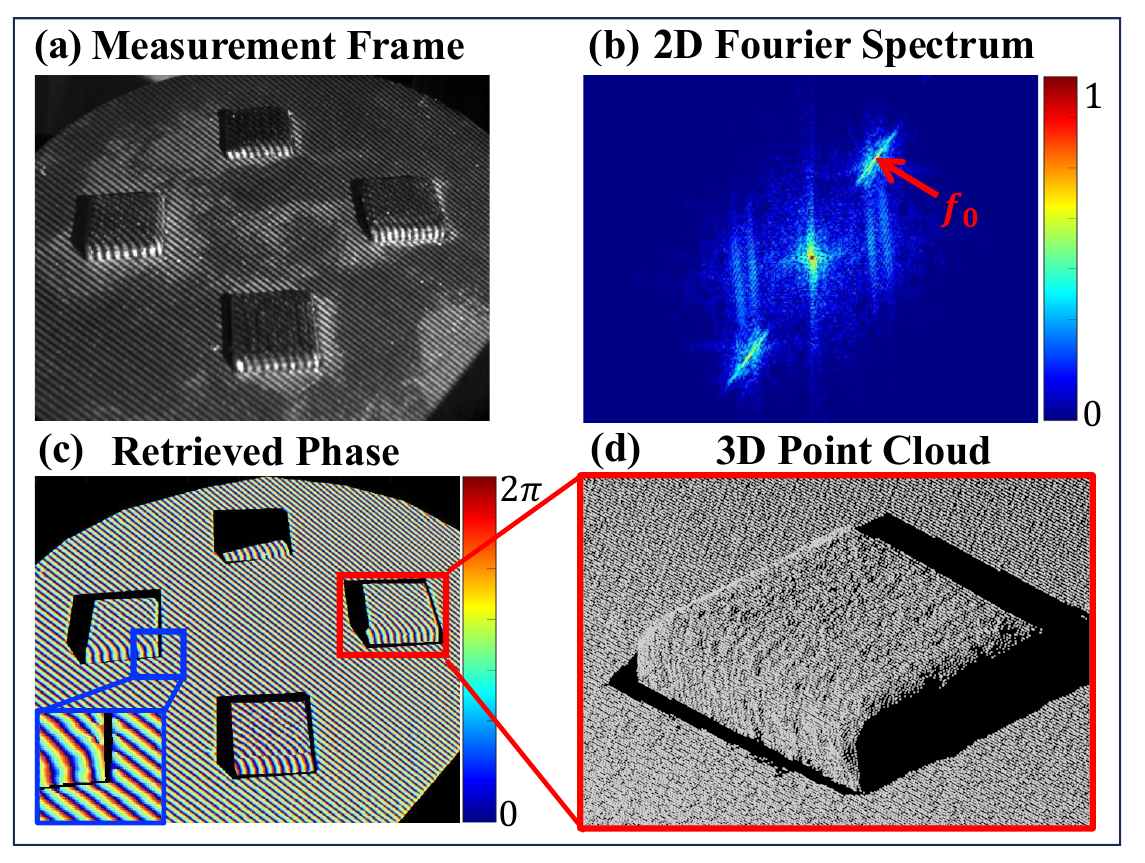}
\caption{Single-shot FPP (FTP) 3D measurement pipeline: (a) Raw captured fringe image. (b) 2D Fourier spectrum. (c) Obtained wrapped phase map. (d) Final 3D reconstruction.}
\label{fig:FrequencySpectrum}
\end{figure}

Traditional FPP techniques rely on the projection of a temporal series of phase-stepped patterns, meaning that a temporal sequence of multiple camera images needs to be captured to calculate one single-shot 3D image. As discussed, this multi-shot acquisition can become problematic for dynamic or unstable measurement environments in DED, where even slight motion or thermal fluctuations between frames can lead to reconstruction artifacts. To ensure motion robustness of each captured 3D view, we adopt a specific flavor of FPP (based on Fourier Transform Profilometry (FTP) \cite{takeda1983fourier,su2001fourier}), which can be performed in a single shot, i.e., it only requires one camera image to calculate one 3D view. 
In FTP, a high-frequency sinusoidal fringe pattern is projected onto the object surface and a single camera image (such as shown in  Fig.~\ref{fig:FrequencySpectrum}(a)) is captured. The recorded sinusoidal intensity distribution can be expressed as
\begin{equation}
I(x,y) = a(x,y) + b(x,y)\cos\!\big(2\pi f_0 x + \varphi(x,y)\big)
\label{eq:ftp_spatial_real}
\vspace{1mm}
\end{equation}

\noindent
where \(a(x,y)\) is bias illumination, \(b(x,y)\) is modulation amplitude, \(f_0\) is carrier frequency, and \(\varphi(x,y)\) is object-encoded phase, which is used to determine the correspondence. We apply a two-dimensional Fourier transform to separate this into a DC term and first-order lobes, as illustrated in Fig.~\ref{fig:FrequencySpectrum}(b)

\begin{equation}
\mathcal{F}\{I(x,y)\} = A(f_x,f_y) + B(f_x - f_0,f_y) + B^*(f_x + f_0,f_y)
\label{eq:ftp_frequency}
\end{equation}
\vspace{1mm}

\noindent 
Figure~\ref{fig:FrequencySpectrum}(b) shows the normalized magnitude of this two-dimensional Fourier spectrum. The horizontal and vertical axes correspond to the spatial-frequency coordinates \(f_x\) and \(f_y\), respectively, with the DC term located at the center. The color represents the normalized spectral magnitude on a scale from 0 to 1. The red arrow marks the selected first-order lobe at the carrier frequency \(f_0\), as represented in Eq.~\ref{eq:ftp_frequency}. These lobes are aligned around the respective carrier frequency and contain a copy of the object spectrum. We use a Hanning band-pass filter \cite{blackman1958measurement} to isolate one first-order lobe, and apply an inverse Fourier transform to obtain the wrapped phase map of our sinusoidal camera image: 

\begin{equation}
\varphi(x,y) = \arg\left[\mathcal{F}^{-1}\{B(f_x - f_0,f_y)\}\right]
\label{eq:wrapped_phase}
\end{equation}
\vspace{1.5mm}

\noindent
shown in Fig.~\ref{fig:FrequencySpectrum}(c). The wrapped phase must then be unwrapped to obtain a continuous phase distribution across the surface, needed to calculate the 3D coordinates. This can be done by a variety of strategies, such as reference markers, a known reference surface like the build plate, or stereo correspondence between multiple views \cite{willomitzer2019single}. For the examples shown in this paper, we use the build plate (which is seen by all cameras) as a reference surface. The so-obtained final phase and correspondence map are then used to calculate the 3D point cloud of the surface (see Fig.~\ref{fig:FrequencySpectrum}(d)) via triangulation (see Fig. \ref{fig:tri_schematic}).

\vspace{3mm}
\subsection{\textcolor{black}{Quantitative evaluation of 3D data precision}}
\label{subsec:precision}

The depth precision of active triangulation measurements is influenced by several system parameters, such as triangulation angle $\theta$, lateral camera resolution or SNR. Understanding the interplay of all involved parameters is key for a physical optimization of the hardware setup~\cite{hausler2022reflections}, meaning that the resulting  tradeoff-space can  be optimized towards a specific goal (e.g., high speed, large FoV, etc.), which is dictated by the targeted application. We refer to \cite{hausler2022reflections, willomitzer2019single} for an overview of involved limits and tradeoffs. 

For a quantitative depth precision evaluation of our proposed FPP system, we performed controlled 3D measurements on objects with well-defined geometries. To emulate the measurement scenario as realistically as possible, we measured a metallic planar object (the build plate) inside the DED chamber. The depth precision of the measured point cloud was then quantified by fitting a low-frequency polynomial surface to the reconstructed point cloud and then calculating the root-mean-square error (RMSE) of the residual point distances between the measured 3D points and the best-fit surface. We performed this procedure for our single-shot FPP (FTP) method described in sec.~\ref{subsec:FTP} and, for comparison, also for the standard multi-shot FPP method. Importantly, as the triangulation angle and other system parameters slightly vary for each FPP sub-sensor, we performed this evaluation for all 4 commonly used FPP sub-sensors in our multi-view system. The evaluated depth precision values $\delta z$ ranged from 37 \textmu m to  44\textmu m\  (mean value $\delta z_{mean} = 40.3 \mu m$ ) for single-shot and 28\textmu m to 48\textmu m (mean value $\delta z_{mean} = 35.5 \mu m$ ) for a multi-shot phase-shifting FPP comparison that we have performed using a total of 12 images per camera (4 phase shifts at 3 different frequencies). This indicates that our system maintains comparable precision to multi-shot FPP while providing substantially improved acquisition speed and robustness against motion artifacts.

\begin{table}[b]

\centering

\caption{Single-shot FPP depth precision $\delta z$ versus surface roughness $R_a$.}

\label{tab:roughness}

\begin{tabular}{lcccccc}

\toprule

$R_a$ ($\mu$m) & 0.05 & 0.10 & 0.20 & 0.40 & 0.80 & 1.60 \\

\midrule

$\delta z$ ($\mu$m) & 26.1 & 30.7 & 44.9 & 37.4 & 39.8 & 38.6 \\

\bottomrule

\end{tabular}

\end{table}

Furthermore, we evaluated the influence of surface roughness on our measurement precision using a commercial Microfinish comparator set (UpAmarket). Multiple flat metallic regions with roughness values ranging from $Ra = 0.05\mu m$ to $1.6\mu m$ were selected and measured with our system in a single frame. The measured depth precision showed a moderate dependence on surface roughness, and ranged from approximately $25 \mu m$ to $45\mu m$ over the tested roughness range. Detailed results are summarized in Table \ref{tab:roughness}. The lateral resolution
of our system is mainly determined by the object-sided pixel pitch, i.e., the lateral distance between two sampling points. This distance was calculated to $110\mu m$, which is significantly larger than the diffraction limited resolution of our optical system, meaning that the lateral resolution can still be improved by reducing the object-sided pixel pitch.
In addition to the described precision evaluations on flat surfaces, we have evaluated the depth precision of a measurement with more complex geometry. Respective measurements and results are discussed in sec. \ref{subsec:insituSS}. 

\vspace{1mm}
\subsection{\textcolor{black}{Multi-view registration and normal-based pointcloud fusion}}
\label{subsec:registrationprocess}
To achieve complete 360° coverage of the object geometry, the partial reconstructions from each FPP sub-system must be accurately registered into a unified coordinate system. This is accomplished through a two-stage registration pipeline consisting of an initial coarse alignment, followed by a fine alignment and view fusion step.

\begin{figure*}[!t]
    \centering
    \includegraphics[width=\linewidth]{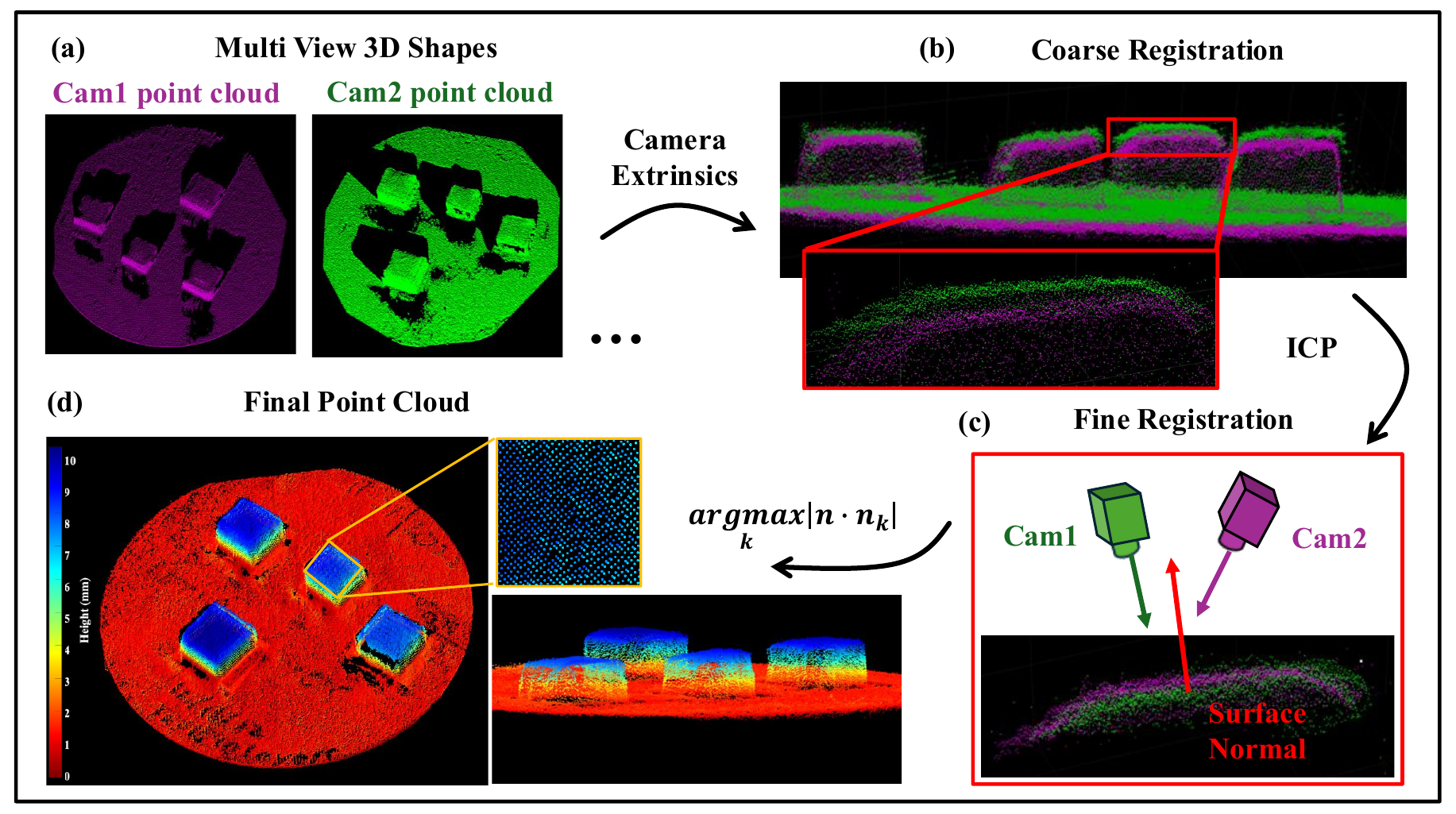}
    \caption{Multi-view point cloud fusion (example with two views): (a) Reconstruct 3D point clouds from two different viewpoints. (b) Perform coarse alignment using calibrated camera extrinsics; a noticeable gap remains between the two point clouds. (c) Apply fine alignment via iterative closest point (ICP) to minimize inter-cloud discrepancies. (d) Compute the angle between the surface normal and each camera’s optical axis, and retain the point cloud from the camera with the smaller angular deviation. The fused reconstruction reveals clear surface ripples on the object. }
    \label{fig:registration_pipeline}
\end{figure*}

In the first stage, all views are transformed into the coordinate frame of Camera 1 using the calibration parameters obtained during system calibration and joint pose optimization (see sec. \ref{subsec:FPPcalibration}). An example point cloud is shown in Fig.~\ref{fig:registration_pipeline}(b). However, due to inevitable remaining calibration inaccuracies, an additional fine registration stage is required. For this, we first identify overlapping regions in the point clouds from the different captured 3D views. These regions are observed across multiple cameras, and can be identified using a combination of build plate visibility information and surface segmentation based on local surface normals calculated from the respective point clouds. Once the overlapping surfaces are identified, we perform an Iterative Closest Point (ICP) \cite{besl1992method,arold2014hand, willomitzer2013flying} alignment between the identified overlap regions. The results before and after the ICP are shown in Fig.~\ref{fig:registration_pipeline}(b) and (c). It can be seen that the different point clouds are now largely overlapping in one surface.

Following the ICP multi-view registration, multiple views from multiple FPP sub-sensors may contribute redundant measurements in overlapping regions. This can pose a challenge for fine high-frequency surface features, such as hatch ripples \cite{GUAN2024200174}: Remaining registration errors can lead to  small translations or rotations between each multi-view point cloud, effectively ``washing out'' fine features. To address this potential issue, we implement a surface normal-based filtering strategy that retains only the most reliable point cloud for each overlap region that is assumed to provide the highest quality data. For each set of overlapping regions, we take the computed set of surface normals, and compare their angles with the position and direction of each camera's optical axis. 

\vspace{-0.8mm}
The quality metric for a given surface patch is then inferred from the relative angle between the mean normal vector $\mathbf{n}$ (see Fig.~\ref{fig:registration_pipeline}(c)) and the camera’s optical axis $\mathbf{n}_k$, defined as the $z$-axis in each camera’s frame, i.e., $\left[0,\,0,\,1\right]^T$. Points observed under more direct (i.e., near-normal) viewing angles are assumed to deliver higher quality  point clouds and are prioritized, as these are less affected by oblique perspective distortion and are more likely to preserve fine geometric features. All other overlapping point clouds are discarded. The final 360° reconstruction retains only points from the “best-viewed” point clouds. 
This view-selective fusion ensures high-fidelity surface representation and avoids local blurring from multi-view redundancy. The result is a complete and geometrically consistent 360° point cloud (Fig.~\ref{fig:registration_pipeline}(d)), constructed from the best portions of each view without introducing blending artifacts. 

\begin{figure}[b!]
    \centering
    \includegraphics[width=1\linewidth]{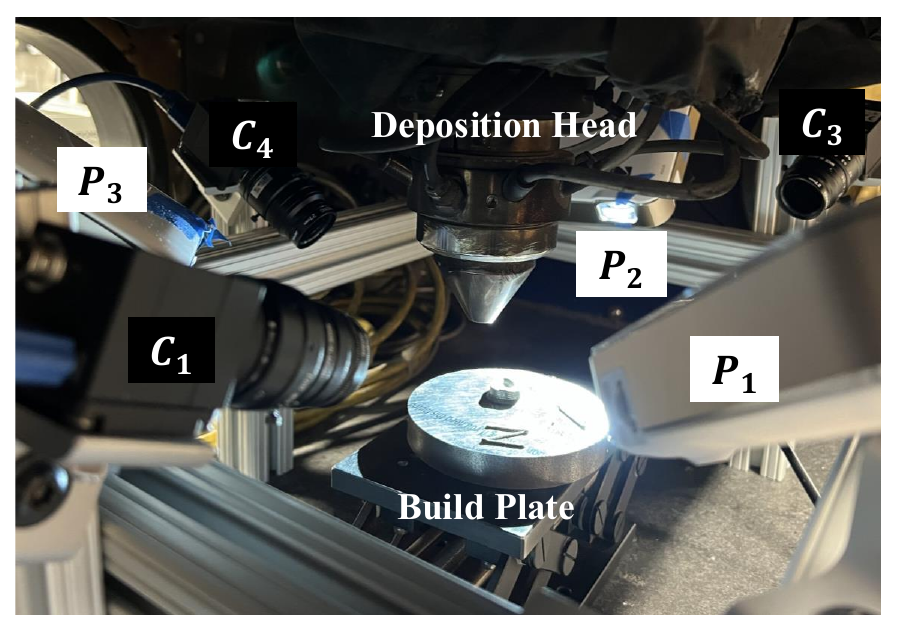}
    \caption{Our 360° 3D metrology system inside the DED chamber.}
    \label{fig:insitu}
\end{figure}

\vspace{1mm}
\section{\textcolor{black}{In-situ Measurement Results and   \\ Evaluation}} 
\label{sec:InssituResults}

To validate the proposed system under realistic conditions, we deployed the full 360° metrology setup inside the functional DED machine described in sec. \ref{subsec:DED}. The multiview system shown in Fig.~\ref{fig:layout} was designed to both fit inside the chamber and avoid moving internal parts. Our system consists of four 2~Megapixel cameras (FLIR BFS-U3-19S4M) equipped with \(8\)~mm objective lenses and three projectors (Kodak Luma 350) arranged around the measurement volume. The projectors are positioned with approximately 120° angular separation, enabling full coverage of the target volume. The cameras are similarly arranged, and an additional fourth camera is present to allow for additional 3D information to be obtained from each projector. Figure~\ref{fig:insitu} shows the complete assembly mounted on a modular aluminum frame and placed inside the DED chamber.

In the current implementation, the in-situ 3D measurements were performed between deposition steps. After the completion of a deposited layer or track segment, the laser and powder feed were turned off, and the deposition head was paused and moved away from the measurement region to avoid optical obstruction of the cameras and projectors. The single-shot FPP system then acquired the surface geometry before the next deposition step was started. Therefore, the measured surface corresponds to the solidified post-deposition morphology, rather than the instantaneous melt-pool geometry during active laser deposition. Any shrinkage or morphology evolution occurring during the short cooling period before image acquisition is included in the measured geometry, which represents the actual surface condition for the subsequent layer. 

The following results demonstrate and evaluate our system’s performance for different imaging scenarios: fast single-shot 3D reconstruction from one FPP sub-system and 360° multi-view reconstruction and registration of a complicated part from 4 FPP sub-systems.

\begin{figure*}[b]
\centering
\includegraphics[width=0.7 \linewidth]{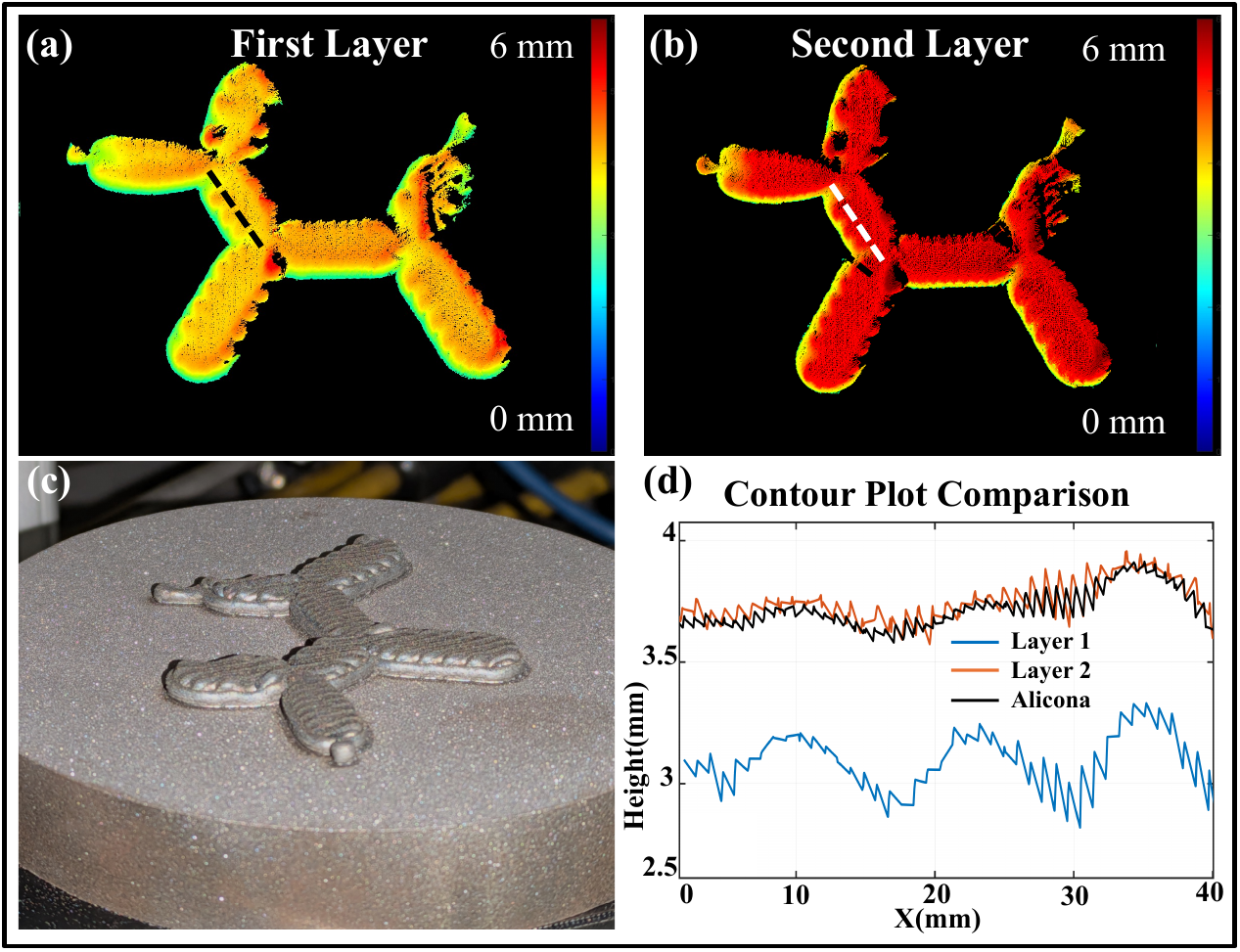}
\caption{In-situ measurement of single-layer DED deposition with fast single-shot FPP. Two layers of a ``balloon dog'' are subsequently printed on a build plate. (a) Height map after printing the first layer. (b) Height map after printing the second layer. (c) Image of the ``balloon dog'' part that is being measured. (d) Height profiles along the black and white dashed lines drawn in both (a) and (b), respectively, shown together with the Alicona reference profile of the final printed state (second layer), extracted along the same cross-section and registered to the measured data. The height difference between the two layers, as well as hatch ripples, can be resolved.}
\label{fig:BalloonDog}
\end{figure*}

\subsection{\textcolor{black}{In-situ single-shot 3D measurements}}
\label{subsec:insituSS}

The first set of experiments demonstrates the sensitivity of our single-shot method (described in sec. \ref{subsec:FTP}) to single-layer DED depositions. We acquire  single-shot FPP measurements  of the metallic ``balloon dog'' print, shown in Fig.~\ref{fig:BalloonDog} at multiple stages of the printing process. The programmed vertical nozzle increment was $700\mu m$ per layer. Figure~\ref{fig:BalloonDog} shows the 3D reconstruction  before (Fig.~\ref{fig:BalloonDog}(a)), and after (Fig.~\ref{fig:BalloonDog}(b))  the deposition of one additional layer, with the laser and powder feed turned off and the deposition head moved away from the field of view. The point cloud overlays clearly reveal localized height changes corresponding to the deposited regions while maintaining stable geometry in surrounding areas. A depth profile drawn across the deposition zone (Fig.~\ref{fig:BalloonDog}(d)) shows a distinct offset between the two layer states. The initial layers are deposited onto a room-temperature substrate with different thermal and material properties from those of the Inconel 718 deposit, which can lead to unstable melt-pool behavior and deviations between the programmed nozzle step and the actual deposited height~\cite{hu2024digital}. The slightly varying  height difference observed in the measured profile is therefore consistent with the expected nominal layer increment, considering the transient instability of the first deposited layers.

We evaluate the depth accuracy of our reconstruction by comparing our measured result to a measurement from an industry-grade focal scanning microscope (Alicona InfiniteFocus G4) for optical metrology applications. The microscope lists a depth resolution of $5\mu m$, which is significantly smaller than our initially evaluated precision values, meaning that the respective microscope measurement can serve as a good approximation of ground truth for our measurement.  After outlier removal, we registered our reconstructed point cloud (27,000 3D points) to the ground truth point cloud from Alicona via ICP fine alignment and calculated the RMSE of our measured point cloud with respect to the ground truth. For the balloon-dog measurement from Fig.~\ref{fig:BalloonDog}(b), the depth accuracy of our measurement calculated with this procedure was evaluated to $64\mu m$. Figure~\ref{fig:BalloonDog}(d) shows the overlaid cross-sectional profiles of the Alicona reference (black) and our measurement (orange). Calculating the difference for the 3D points only shown in these profiles delivers an RMSE of $55\,\mu\mathrm{m}$, slightly lower than the calculated error of the whole point cloud. We emphasize that the calculated accuracy contains the statistical error of our measurement as well as our system’s systematic errors, such as calibration errors. 
For better comparison with our previous analysis shown in sec.~\ref{subsec:precision}, we have additionally estimated the precision of the measurement shown in Fig.~\ref{fig:BalloonDog}(b),   using a similar method we used on the previously measured planar surfaces: After subtraction of the Alicona ground truth data from our data, we fitted a low-frequency polynomial surface to the residual point cloud to estimate the low-frequency systematic error of our measurement. This low-frequency surface was then again subtracted and the RMSE of the remaining point cloud was calculated, which serves as an estimate of the precision of our measurement on the balloon dog part. This precision estimate was calculated to $55\mu m$, which is in good agreement with the precision values obtained in sec.~\ref{subsec:precision}.

\begin{figure*}[b!]
\centering
\includegraphics[width=0.7\linewidth]{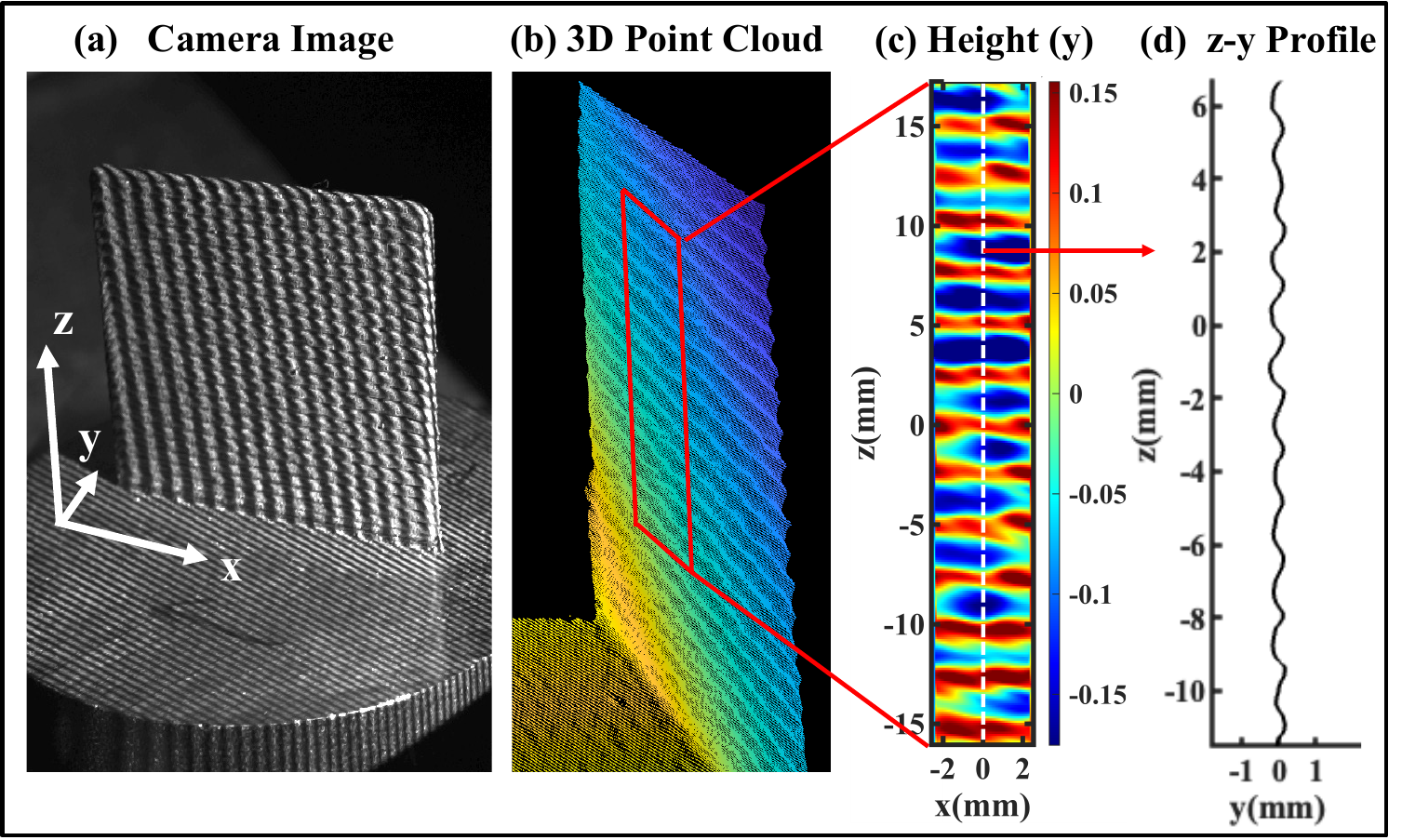}
\caption{In-situ single-shot FPP measurement of a tall DED part. (a) Camera image. (b) Evaluated 3d point cloud, showing the side of the part. Different deposition layers can be clearly seen. (c) Height map of the side profile. (d) z-y profile plot to visualize deposition layers.  }
\label{fig:MultiLayer}
\end{figure*}

In an additional quantitative analysis, we further verified that the ripples visible in our surface reconstruction of Fig.~\ref{fig:BalloonDog} (d) indeed correspond to real DED printing process-induced features rather than being introduced by reconstruction artifacts. For this we performed a matched Fourier peak analysis on both the Alicona ground truth measurement and our registered FPP reconstruction. 
After removing the low-frequency shape component, the Alicona ground truth measurement exhibited a dominant spatial ripple period of approximately $816-819\mu m$ , while our FPP reconstruction showed a dominant ripple period of approximately $801-816\mu m$. The close results indicate that the observed surface ripples are faithfully captured by our system. Furthermore, both measured periods are within the printing tolerance of the nominal hatch spacing of~$730\mu m$.

In a second experiment, we further validate our method on a taller DED-built part with many accumulated layers (Fig.~\ref{fig:MultiLayer}). The  3D point cloud (Fig.~\ref{fig:MultiLayer}(b)) taken from the side of the part shows that our single-shot method nicely captures different deposition layers. A height map (Fig.~\ref{fig:MultiLayer}(c)) and line profile along the side surface (Fig.~\ref{fig:MultiLayer}(d)) reveal the  structure of the deposited layers, where each hump corresponds to one DED pass. These results demonstrate that the system not only detects subtle single-layer increments but also resolves stacked layer profiles in taller builds with high fidelity. 

We emphasize again that the results shown in this subsection present  single-shot measurements from one FPP sub-system, and no registration of different 3D point clouds has been applied. The total time (including acquisition and time to evaluate the 3D model) is below 0.5 seconds. This means that, even without further algorithmic optimization, our prototype system is already able to continuously monitor fast 3D printing processes in specific regions of the object on-the-fly, without requiring lengthy interruptions of the printing process during the measurement.

The time improvement becomes obvious when comparing our method to conventional multi-shot phase-shifting FPP implementations. The respective techniques typically need to capture four phase-shifted fringe images for phase retrieval using at least two fringe frequencies for phase unwrapping, resulting in at least eight captured images for a complete 3D reconstruction in one FPP sub-system. Under the same camera operating conditions used in this work ($16Hz$ frame rate), such an implementation would require approximately $0.5s$ for image acquisition alone (excluding computing time), while our approach finishes the whole process in less than $0.5s$. We also emphasize that, beyond the reduction in the number of required images, the proposed single-shot framework is not fundamentally constrained by the camera frame rate anymore. Since the entire 3D geometry is recovered from a single captured image, the achievable speed of a single view 3D measurement  is  limited only by the exposure time of that image. Consequently, even faster measurements may be achieved through shorter exposure times enabled by brighter illumination or stroboscopic acquisition.

\begin{figure*}[t]
\centering
\includegraphics[width=0.9\linewidth]{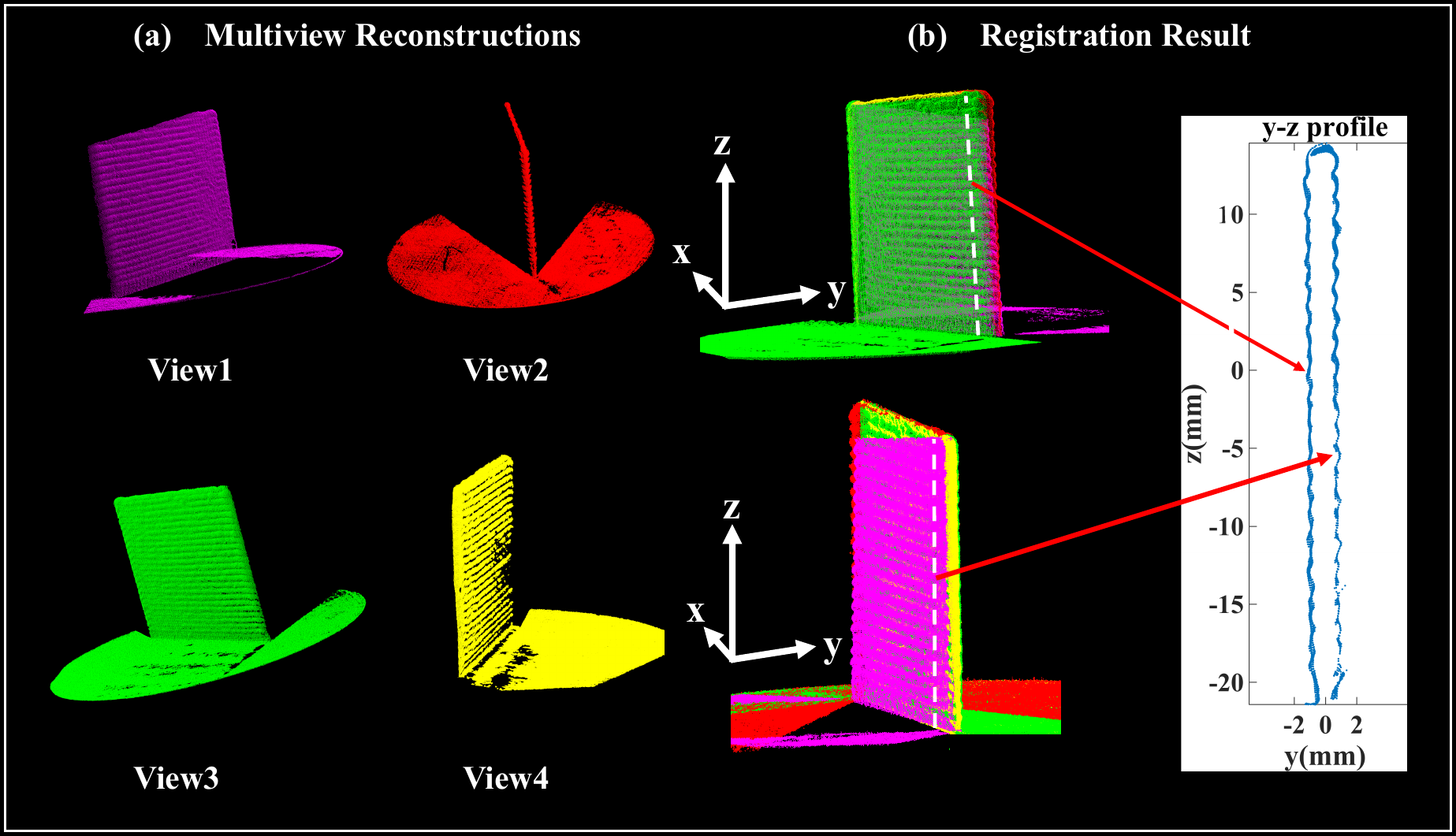}
\caption{In-situ 360° 3D measurement of a printed thin wall part. (a) Individual 3D point clouds from all the FPP sub-sensors used in this measurement. (b) Two views of the final registration result. Using the baseplate as an XY plane, a height profile is plotted along the white dashed lines in the final point cloud, showing a representative cross section of our surface reconstruction. The thin wall geometry presents a unique challenge to conventional ICP registration as the two closely spaced wall surfaces could be registered together. Due to our surface normal-based detection of overlap regions in the individual point clouds, our registration algorithm is robust to this error and can reconstruct the thin wall object with its true thickness in high quality.}
\label{fig:wallpart}
\end{figure*}

\subsection{In-situ 360° multi-view measurement of a complicated part}
\label{subsec:wallpart}
To evaluate the robustness of our registration algorithm, we performed a full in-situ  360° measurement of the same metallic part (see Fig.~\ref{fig:MultiLayer}(a)) whose single point cloud view was evaluated in sec.~\ref{subsec:insituSS}. The object consists of a single printed thin wall. This geometry of two closely spaced planar surfaces separated only by the wall thickness presents a particularly challenging task for multi-view registration algorithms, and a Iterative Closest Point (ICP) alignment is highly prone to error in such cases because ICP minimizes point-to-point Euclidean distances; the standard approach tends to collapse two nearly parallel planes into a single surface, effectively erasing the true wall thickness. The result would be an incorrect 3D model where opposing wall surfaces are spuriously merged.

For the 360° measurement shown in  Fig.~\ref{fig:wallpart}, we collect four individual point clouds from four FPP sub-sensors. Figure~\ref{fig:wallpart}(a) shows the four individual point clouds prior to registration. Each point cloud contains only part of the structure, and, as the figure illustrates, the thin wall produces three planar point clusters (views 1,3, and 4) that lie in close proximity. As discussed, our developed registration algorithm addresses this challenge by exploiting surface normal information to determine valid overlap regions (see sec.~\ref{subsec:registrationprocess}). In the specific case of thin walls, surfaces viewed by opposing cameras whose normals are nearly antiparallel are explicitly rejected as candidate overlaps and the respective parts of the point clouds are not ICP registered to each other. The benefit of this strategy is evident in Fig.~\ref{fig:wallpart}(b). After registration, the four partial reconstructions are fused into a unified model that preserves the distinct geometry of each wall surface. A representative cross-section (Fig.~\ref{fig:wallpart}(b) right) taken along the dashed lines in the figure confirms that the reconstructed profile maintains the true thickness of the wall, with no spurious collapse into a single plane. To quantify the multi-view registration error, we evaluated  residual distances between overlapping sub-sensor reconstructions after calibrated alignment and ICP refinement. For each overlapping pair, we removed the statistical noise from each respective point cloud by calculating a best-fit surface, and eventually quantified the registration error by calculating the RMS of the residual distance between both best-fit surfaces. The resulting registration error was \(\sigma_{\mathrm{reg}}= 20 \mu m\).

In conclusion, the  shown  experiment demonstrates that our normal-based registration produces stable and accurate results even in scenarios that are challenging for ICP-based methods. More broadly, it highlights the ability of the approach to preserve fine structures and prevent topological errors in multi-view 3D reconstruction of multilayer DED structures.

\section*{Discussion and Outlook}
\label{sec:discussion}

We presented a fast, precise, and accurate 360° 3D metrology system for in-situ DED monitoring. The presented results demonstrate that our system enables single-shot, high-precision 360° metrology of shiny metallic printed parts inside the DED chamber. By combining single-shot FPP (FTP) with cross-polarization filtering, the approach effectively suppresses specular reflections and captures surface geometry with sufficient sensitivity to resolve both single-layer changes and stacked layer profiles.

The depth precision of our system has been evaluated from all performed quantitative measurements to  $\delta z < 55 \mu m$, and the reconstruction accuracy evaluated on a complex printed part has been quantified to  $64 \mu m$. Our multi-view registration further extends coverage, allowing partial reconstructions to be fused into a complete point cloud while mitigating residual specular saturation through complementary viewpoints. We see our prototype system as a first step towards dynamic, in-situ DED monitoring and inspection, where motion and thermal instabilities make approaches relying on long measurement sequences less practical.

In our experiments, we did not observe significant degradation of the reconstructed geometry caused by thermal air turbulence during the DED process. This is primarily because the measurements were acquired between deposition steps rather than during active laser-material interaction. Thermal turbulence in DED is mainly associated with the melt pool, vapor/plasma plume, and local gas heating during laser exposure. A previous study \cite{wolff2019situ} has shown that the melt pool can solidify within milliseconds after laser shutoff due to the extremely fast cooling rates in DED processes. However, even if stronger thermal turbulence were present or if a measurement during the printing process is desired, the proposed single-shot FTP framework is inherently robust. The rapid, small path changes from turbulence are integrated over the exposure time and lead to a reduction in fringe contrast. As we recover the phase information from a single captured image through frequency-domain analysis, reliable phase retrieval is still possible as long as the dominant carrier frequency component remains detectable.

Our system, however, is not without limitations: although each FPP sub-sensor captures the object surface in a single shot, a multi-view measurement that uses multiple projectors still relies on a short image sequence that cycles through all used projectors. While we accept this limitation for our first proof-of-principle prototype, future versions of our system could address this limitation by, e.g., color projection and filtering or other multiplexing schemes to achieve fully single-shot  multi-view measurements. Moreover, as described in sec.~\ref{sec:InssituResults}, the in-situ measurements are acquired between deposition steps, after the deposition head has moved away and the deposited surface has briefly cooled. The presented system therefore provides layer-wise monitoring rather than truly continuous measurements during active deposition. In addition, the limited number of fixed viewpoints can lead to incomplete spatial coverage: steep sidewalls, undercuts, and self-occluding features may remain unobserved, and certain part geometries would therefore require additional sensor positions for complete coverage.

Our future work will focus on the further refinement of the approach, including system timing and acquisition speed as well as the effective processing and display of the captured data: since the proposed framework provides layer-wise 360° surface measurements throughout the fabrication process, the sequentially acquired point clouds could be accumulated into a volumetric representation of the fabricated part. Such information may enable future capabilities including internal defect localization, cavity tracking, and process-aware digital reconstruction without destructive testing.  

We hope that our framework will be broadly adopted in the future as a general approach for high-speed in-process 3D metrology in industrial environments. Although DED is used in this work as a representative and particularly challenging application scenario, the approach is well suited for potential future applications that involve other reflective metallic surfaces and extend beyond DED to a broad range of manufacturing processes, including additive manufacturing, CNC machining, metal forming, stamping, and other industrial operations requiring rapid geometric inspection and process monitoring.

\section*{Acknowledgements}
This research was supported by the National Science Foundation under Grant number CMMI-2216298. The authors thank Ruijing Zha from the Advanced Manufacturing Processes Laboratory (AMPL) for helpful discussions on the mechanical design of the prototype and its integration into the Additive Rapid Prototyping Instrument (ARPI). A preprint version of this manuscript was submitted to arXiv on September 5th, 2025~\cite{taylor2025fast}.

\section*{References}
\bibliography{bibliography}

\end{document}